\documentclass[%
reprint,
superscriptaddress,
altaffilletter,
showpacs,
showkeys,
amsmath, amssymb, aps,prl, floatfix,
]{revtex4-1}
\usepackage{graphicx}
\usepackage{dcolumn}
\usepackage{bm}
\usepackage{comment}
\usepackage[usenames,dvipsnames]{color}
\usepackage[dvipsnames]{xcolor}
\usepackage[normalem]{ulem}
\usepackage[symbol]{footmisc}
\usepackage{xspace}
\usepackage{ulem}
\usepackage{hyperref}
\hypersetup{colorlinks = true, allcolors = blue}
\usepackage[mathlines]{lineno}
\usepackage{multirow}
\usepackage[version=3]{mhchem}
\usepackage{siunitx}
\usepackage[flushleft]{threeparttable}

\DeclareGraphicsExtensions{.pdf,.png,.jpg}

\newcommand{\orcid}[1]{\href{https://orcid.org/#1}{\includegraphics[width=8pt]{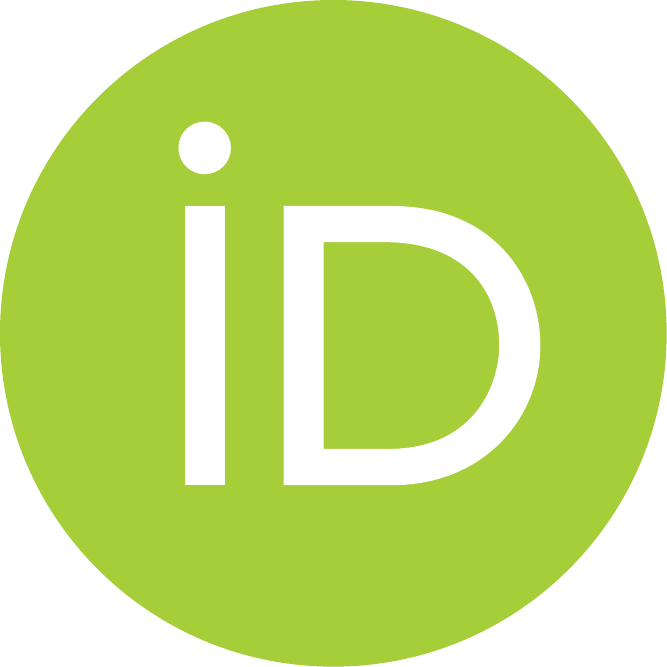}}}

\begin{document}

\title{Simultaneous Measurement of Half-Life and Spectral Shape of \texorpdfstring{$^{115}$}{115-}In \texorpdfstring{$\beta$}{beta}-decay with an Indium Iodide Cryogenic Calorimeter}

\newcommand{\lngs}{\affiliation{INFN - Laboratori Nazionali del Gran Sasso, I-67100 Assergi (AQ) - Italy}}
\newcommand{\infnge}{\affiliation{INFN - Sezione di Genova, I-16146 Genova - Italy}}
\newcommand{\unige}{\affiliation{Dipartimento di Fisica, Università di Genova, I-16146 Genova - Italy}}
\newcommand{\infnmib}{\affiliation{INFN - Sezione di Milano - Bicocca, I-20126 Milano - Italy}}
\newcommand{\unimib}{\affiliation{Dipartimento di Fisica, Università di Milano - Bicocca, I-20126 Milano - Italy}}
\newcommand{\gssi}{\affiliation{Gran Sasso Science Institute, 67100, L'Aquila - Italy}}
\newcommand{\JYU}{\affiliation{University of Jyv\"askyl\"a, Department of Physics, P. O. Box 35 (YFL), FI-40014, Finland}}
\newcommand{\FIER}{\affiliation{Finnish Institute for Educational Research, P.O.Box 35 FI-40014 University of  Jyv\"askyl\"a - Finland}}
\newcommand{\CTP}{\affiliation{Center for Theoretical Physics, Sloane Physics Laboratory, Yale University, New Haven, Connecticut 06520-8120 - USA}}
\newcommand{\QUEEN}{\affiliation{Department of Physics, Engineering Physics and Astronomy, Queen's University Kingston, Ontario, K7L 3N6 Kingston, Canada}}
\newcommand{\PGPI}{\affiliation{Prokhorov General Physics Institute of the Russian Academy of Sciences 119991, Moscow, 38 Vavilov str., Russia}}
\newcommand{\NRI}{\affiliation{Natural Resources Institute Finland, Yliopistokatu 6B, FI-80100 Joensuu, Finland}}
\newcommand{\CIFRA}{\affiliation{International Centre for Advanced Training and Research in Physics, 077125 Bucharest-Magurele, Romania}}

\author{L. Pagnanini}\gssi\lngs\QUEEN 
\author{G. Benato}\gssi\lngs
\author{P. Carniti}\unimib\infnmib
\author{E. Celi}\gssi\lngs
\author{D. Chiesa}\unimib\infnmib
\author{J. Corbett}\QUEEN
\author{I. Dafinei}\gssi
\author{S. Di Domizio}\unige\infnge
\author{P. Di Stefano}\QUEEN
\author{S. Ghislandi}\email[Corresponding author: ]{stefano.ghislandi@gssi.it}\gssi\lngs 
\author{C. Gotti}\infnmib
\author{D. L. Helis}\email[Corresponding author: ]{dounia.helis@lngs.infn.it}\gssi\lngs 
\author{R. Knobel}\QUEEN 
\author{J. Kostensalo}\NRI
\author{J. Kotila}\JYU\FIER\CTP
\author{S. Nagorny}\QUEEN
\author{G. Pessina}\infnmib
\author{S. Pirro}\lngs
\author{S. Pozzi}\infnmib
\author{A. Puiu}\lngs
\author{S. Quitadamo}\gssi\lngs
\author{M. Sisti}\infnmib
\author{J. Suhonen}\JYU\CIFRA
\author{S. Kuznetsov}\PGPI

\date{\today}

\begin{abstract}
Current bounds on neutrino Majorana mass are affected by significant uncertainties in the nuclear calculations for neutrinoless double-beta decay. A key issue for a data-driven improvement of the nuclear theory is the actual value of the axial coupling constant $g_A$, which can be investigated through forbidden $\beta$-decays. We present the first measurement of 4$^{th}$-forbidden $\beta$-decay of $^{115}$In with a cryogenic calorimeter based on Indium Iodide.
Exploiting the enhanced spectral shape method for the first time to this isotope, our study accurately determines simultaneously spectral shape, $g_A$, and half-life. The Interacting Shell Model, which best fits our data, indicates a half-life for this decay at $T_{1/2}=(5.26\pm0.06) \times 10^{14}$\,yr.
\end{abstract}

\pacs{07.20.Mc, 23.40.-s, 21.10.Tg, 27.50.+e}
\keywords{forbidden beta decay, spectral shape}
\maketitle
{\it Introduction.} The search for neutrinoless double-beta decay ($0\nu\beta\beta$) is a crucial part of our quest to understand the deepest mysteries of the universe~\cite{Agostini:2022zub}. The observation of this phenomenon would require a paradigm shift from the standard model of elementary particles and would reshape our understanding of the fundamental building blocks of matter. $0\nu\beta\beta$ is an extremely rare process where two neutrons in the nucleus are simultaneously transformed into protons, with the emission of just two electrons in the final state. If we observe this process, it would indicate that neutrinos are Majorana particles, which means they are their own antiparticles. The half-life of this process ($T^{0\nu}_{1/2}$) could provide insights into the absolute mass scale of neutrinos, which is still an unsolved issue in particle physics. Moreover, $0\nu\beta\beta$ is a lepton-number-violating transition, and its observation would support exciting theoretical frameworks in which leptons played a crucial role in creating the matter/antimatter asymmetry in the universe~\cite{Buchmuller:2005eh,Canetti:2012zc}. The next-generation experiments in this field are designed to approach half-lives of the order of $10^{27}$--$10^{28}$ yr. The current most stringent limit is set on $^{136}$Xe by KamLAND-Zen at $2.3\times10^{26}$\,yr at 90\% C.L.~\cite{KamLAND-Zen:2022tow}. This limit can be converted into a constraint on the effective Majorana mass ($m_{\beta\beta}$), which is the new-physics parameter governing $0\nu\beta\beta$, obtaining $m_{\beta\beta} < 36$--$156$\,meV. 
It is notable that a single value of $T^{0\nu}_{1/2}$ can correspond to a wide range for $m_{\beta\beta}$. This is due to the uncertainty of a factor of 3 affecting the Nuclear Matrix Element (NME) calculation for $0\nu\beta\beta$ within different nuclear models. This uncertainty not only limits the conversion of the half-life into $m_{\beta\beta}$ in case of discovery, but also severely restricts the selection of relevant isotopes for the next-to-next generation of experiments. It is well-known that isotopes with lower Q-values are disfavoured by lower phase space factors, but this precious information could be misleading if the NME landscape is unclear. Therefore, a data-driven improvement of nuclear models is essential to ensure that theoretical and experimental efforts in the $0\nu\beta\beta$ sector are not nullified.
Some of the data and physical processes that could help clarify the puzzle include double-charge exchange reactions~\cite{NUMEN:2021kme,NUMEN:2021ezg}, ordinary muon capture~\cite{Hashim:2023zwf,Gimeno:2023dxx}, two neutrino double beta decay~\cite{Barea:2013bz,CUPID-Mo:2023lru}, and forbidden $\beta$-decay~\cite{Ejiri:2019ezh}. In particular, the latter is very interesting to investigate the origins of the quenching of the axial coupling constant ($g_A$). Indeed, the shape of the forbidden non-unique $\beta$-decay spectrum shows a strong dependence on the value of $g_{A}$~\cite{PhysRevC.95.024327}. In this context, several isotopes have been studied such as $^{113}$Cd~\cite{Belli:2007zza,Bodenstein-Dresler:2018dwb,Kostensalo2021}, $^{99}$Tc~\cite{Paulsen:2023qnm}, and $^{115}$In~\cite{PhysRevLett.129.232502} using the so-called Spectral Shape Method (SSM)~\cite{Haaranen:2016rzs}. This theoretical framework matches with high precision the spectral shape of experimental data. However, the simultaneous prediction of the decay half-life is often far from being compatible with the measured values. Improvements of the models in this direction have been done during the last years within the so-called enhanced SSM theory~\cite{Behrens1982,PhysRevC.101.064304,EPJA.57.225}, 
where the small relativistic NME (sNME) enters as an additional parameter able to adjust the spectrum to predict the half-life. The theoretical values of the sNME predicted under the Conserved Vector Current (CVC) hypothesis are reported in Tab.~\ref{tab:sNME_CVC}.

In this letter, we present the first application of the enhanced SSM on \ce{^{115}In}. The measurement has been performed with a cryogenic calorimeter based on Indium Iodide (InI) crystal in the framework of the ACCESS (Array of Cryogenic Calorimeters to Evaluate Spectral Shape) project~\cite{ACCESS:2023gdy}.

\begin{table}[t!]
    \centering
    \setlength{\tabcolsep}{6pt}
    \begin{tabular}{r|l}
         {\it Model} & {\it sNME} \\[2pt]
         \hline 
         \\[\dimexpr-\normalbaselineskip+2pt]
         ISM & 6.01 \\
         MQPM & 10.25 \\
         IBFM-2 & 2.53 \\
    \end{tabular}
    \caption{Values of the sNME calculated in the framework of the three nuclear models adopted in this work (ISM - Interacting Shell Model, MQPM - Microscopic Quasi-Particle Phonon Model and IBFM-2  - Interacting Boson-Fermion Model) under the Conserved Vector Current (CVC) hypothesis.}
    \label{tab:sNME_CVC}
\end{table}

\begin{figure}[t!]
    \centering
    \includegraphics[width=0.33\textwidth]{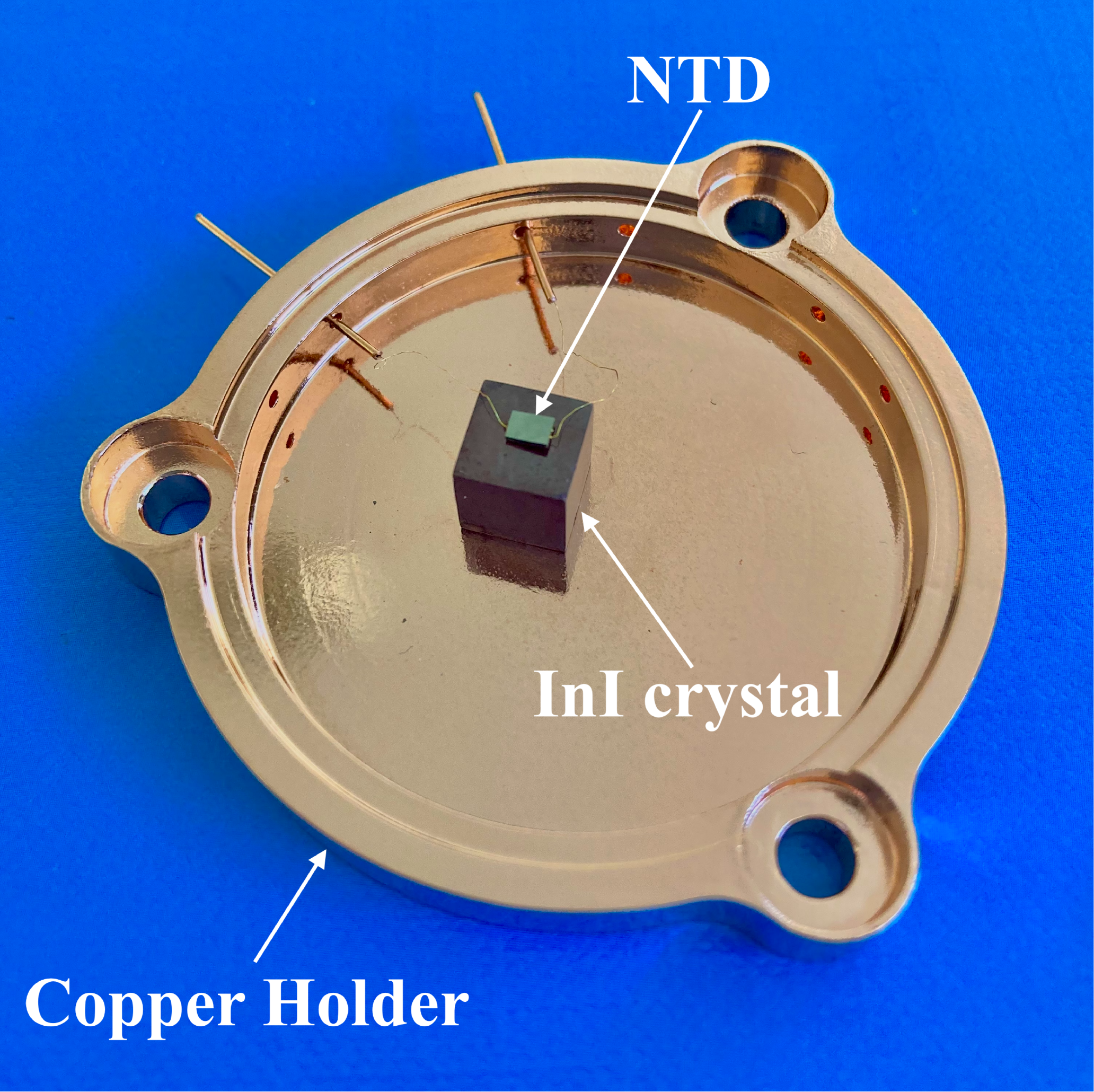}
    \caption{Experimental setup used to measure the Indium Iodide (InI) crystal as cryogenic calorimeter at LNGS. The crystal is equipped with a Neutron Transmutation Doped (NTD) germanium sensor and rests on the copper holder that is connected to the lowest temperature stage (16 mK) of the cryostat utilizing a double-stage vibration damping system. The crystal is thermally linked through the gold wires for the signal readout.}
    \label{fig:in_setup}
\end{figure}

{\it Detector setup.} Following the design principles outlined in Ref.~\cite{ACCESS:2023gdy}, we conduct measurements on a 7$\times$7$\times$7\,mm$^{3}$ Indium Iodide (InI) crystal. The crystal has a mass $m_{\text{InI}}$ of $(1.91 \pm 0.01)$\,g and it is equipped with a Neutron Transmutation Doped (NTD) germanium thermistor (3$\times$2$\times$0.5 mm$^{3}$) to record particle interactions within its lattice. The detector, as shown in Fig.~\ref{fig:in_setup}, rests on a copper holder that is directly connected to the mixing chamber of the CUPID R\&D dilution refrigerator~\cite{Azzolini:2018tum} installed in Hall A of the Laboratori Nazionali del Gran Sasso (LNGS), Italy.
The setup is cooled down to approximately 16\,mK. During the whole data-taking, we use a thoriated wire as \ce{^{232}Th} permanent calibration source mounted close to the detector. Periodic calibrations with external sources would have been more difficult due to the small size of the crystal and the presence of a lead shield in the cryostat itself. The front-end electronics consists of an amplification stage, a six-pole anti-aliasing active Bessel filter, and an 18-bit ADC board~\cite{Arnaboldi:2015wvc,Arnaboldi:2017aek}. The data stream is digitized at a frequency of 2\,kHz and stored on disk in NTuples using a ROOT-based software framework~\cite{DiDomizio:2018ldc}. An online software derivative trigger, incorporating a channel-dependent threshold, flags noise and signal events.

{\it Data analysis.} The offline analysis of the data stream involves calculating several variables for each triggered signal. These variables include the number of triggers in the acquisition window, the slope of the baseline (pre-trigger of the pulse in the acquisition window), the rise time and decay time of the pulses. We exploit these quantities to construct average templates of noise and signal events, needed to apply the Optimum Filter~\cite{Gatti:1986cw}. We adopt this technique to estimate the amplitude of each triggered event by maximizing the signal-to-noise ratio~\cite{Azzolini:2018yye}.
We then perform a stabilization of the detector thermal gain using the 238.6\,keV $\gamma$-ray line from the \ce{^{232}Th} source~\cite{Alduino_2016}.
Consequently, we calibrate in energy the stabilized spectrum using the most prominent peaks visible in the data. We observe an energy resolution of 3.9\,keV (FWHM) at 238.6\,keV.
The detection threshold, defined as 5 times the baseline root-mean-square, is estimated to be 3.4\,keV.
The criteria for selecting the events are based on rejecting noisy acquisition time intervals and windows that have more than one triggered pulse, which is commonly known as a distinguishable pile-up. Additionally, we apply pulse-shape cuts requiring a constant selection efficiency as a function of the energy. This is mandatory to avoid any possible distortion in the spectral shape due to analysis. The overall analysis cut efficiency is $\epsilon = (52.2\,\pm 0.3)$\%, where the distinguishable pile-up cut dominates.

{\it Data Modeling and Spectral Fit.} The study of the \ce{^{115}In} $\beta$-decay shape and the estimation of its half-life can be achieved through a background decomposition of the collected data. For this purpose, the geometry of the experimental setup is implemented into a Geant4-based~\cite{GEANT4:2002zbu} simulation.
In the following, we define as \emph{signal} the $\beta$-spectrum of the \ce{^{115}In} ($Q_{\beta}= 497.489(10)$~keV~\cite{BLACHOT20122391}), and as \emph{background} all the remaining contributions required to explain the energy spectrum measured by the \ce{InI} crystal.
For the signal, we generate electrons with energy sampled from spectra templates based on three different theoretical frameworks: Interacting Shell Model (ISM)~\cite{RevModPhys.77.427}, Microscopic Quasi-Particle Phonon Model (MQPM)~\cite{Toivanen:1998zz} and Interacting Boson-Fermion Model (IBFM-2)~\cite{iachello_isacker_1991}. 
These templates are calculated for fixed values of $g_{A}$ and sNME, which vary in the range [0.60, 1.39] and [-5.9, 5.9] and with steps of 0.01 and 0.1, respectively.
If needed, we use linear spline interpolation to increase the template fine structure. 
The most prominent background comes from the thoriated wire used as a calibration source. 
In order to account for a possible breaking of the secular equilibrium, we separately simulate the partial decay chain from \ce{^{232}Th} to \ce{^{228}Ac} and the remaining one starting from \ce{^{228}Th}.
Any other potential background contribution, whether from the crystal or the cryogenic setup~\cite{CUPID:2023wyy}, is smaller than the statistical uncertainty associated to the bin counts.
This is consistent with the absence of any other features in the spectrum.
The Monte Carlo simulations undergo a post-processing step that takes into account the effects of un-resolvable pile-up and finite energy resolution.
The fit is performed in the energy range of [80, 800]~keV. At lower energies, the data reconstruction is not satisfactory, while at higher energies the statistics is scarce. A uniform binning of 10\,keV is chosen to avoid systematic effects due to the peak line-shape.

We assume the number of counts in each bin to follow the Poisson probability distribution $Pois(n,\nu)$, where $n$ is the number of observed events, and $\nu$ is the expected number of counts. $\nu$ consists of a linear combination of the signal template $S(g_{A},\text{sNME})$ and background simulations $B_{j}$. 
We introduce the normalization factors $N_{S}$ and $N_{B,j}$, that are proportional to the half-life of \ce{^{115}In} ($T_{1/2}$) and to the activities of the background components, respectively.
The half-life can be expressed as
 \begin{equation}
     T_{1/2} = \frac{\mathrm{ln}(2) \cdot t \cdot m_{\mathrm{InI}} \cdot N_{A} \cdot i.a.(\ce{^{115}In}) \cdot \epsilon }{ M_{\mathrm{InI}} \cdot N_{S} \cdot N_{MC} }
 \end{equation}
 where $t = 128.8$\,h is the measurement time, $N_{A}$ is the Avogadro constant, $i.a.(\ce{^{115}In}) = (95.719 \pm 0.052)$\%~\cite{In115ia} is the natural isotopic abundance of \ce{^{115}In}, $M_{\mathrm{InI}}$ is the molar mass of InI, and $N_{MC}$ the number of simulated $\beta$ decays. The expected number of events in the $i$-th bin is
 \begin{equation}
     \nu_{i} = N_{S}(T_{1/2})\cdot S(g_{A},\text{sNME})_{i} + \sum_{j=1,2}N_{B,j}\cdot(B_{j})_{i}
 \end{equation}
where $j$ identifies the \ce{^{232}Th} and \ce{^{228}Th} contributions from the calibration source. The likelihood can be therefore written as
 \begin{equation}
    \label{eq:TotalLikelihood}
     \mathcal{L}(\textbf{data} \,\,|\,\, T_{1/2},g_A,\textsc{sNME},N_{B,j}) = \prod_{i} Pois(n_{i},\nu_{i}).
 \end{equation} 
The fit uses five free parameters, with three continuous, $T_{1/2}$ and $N_{B,j}(j=1,2)$, and the two remaining are discrete, namely, $g_{A}$ and sNME. The discrete parameters identify the theoretical template to be picked at every step of the Markov-Chain Monte Carlo used for the posterior sampling. We assume uniform prior probability distributions for all these parameters. Additionally, we introduce the analysis cut efficiency $\epsilon$ with a Gaussian prior probability distribution.
We use the Bayesian Analysis Toolkit (BAT)~\cite{Schulz:2020ebm} to perform the statistical inference as well as the posterior sampling and marginalization.
For each fit, we quote the median of the marginalized posterior as an estimator of the best value of the parameter at issue. The interval defined by [16, 84]\% quantiles is used to evaluate the uncertainty.
 
 \begin{figure}[t]
  \centering
  \includegraphics[width=1.\columnwidth]{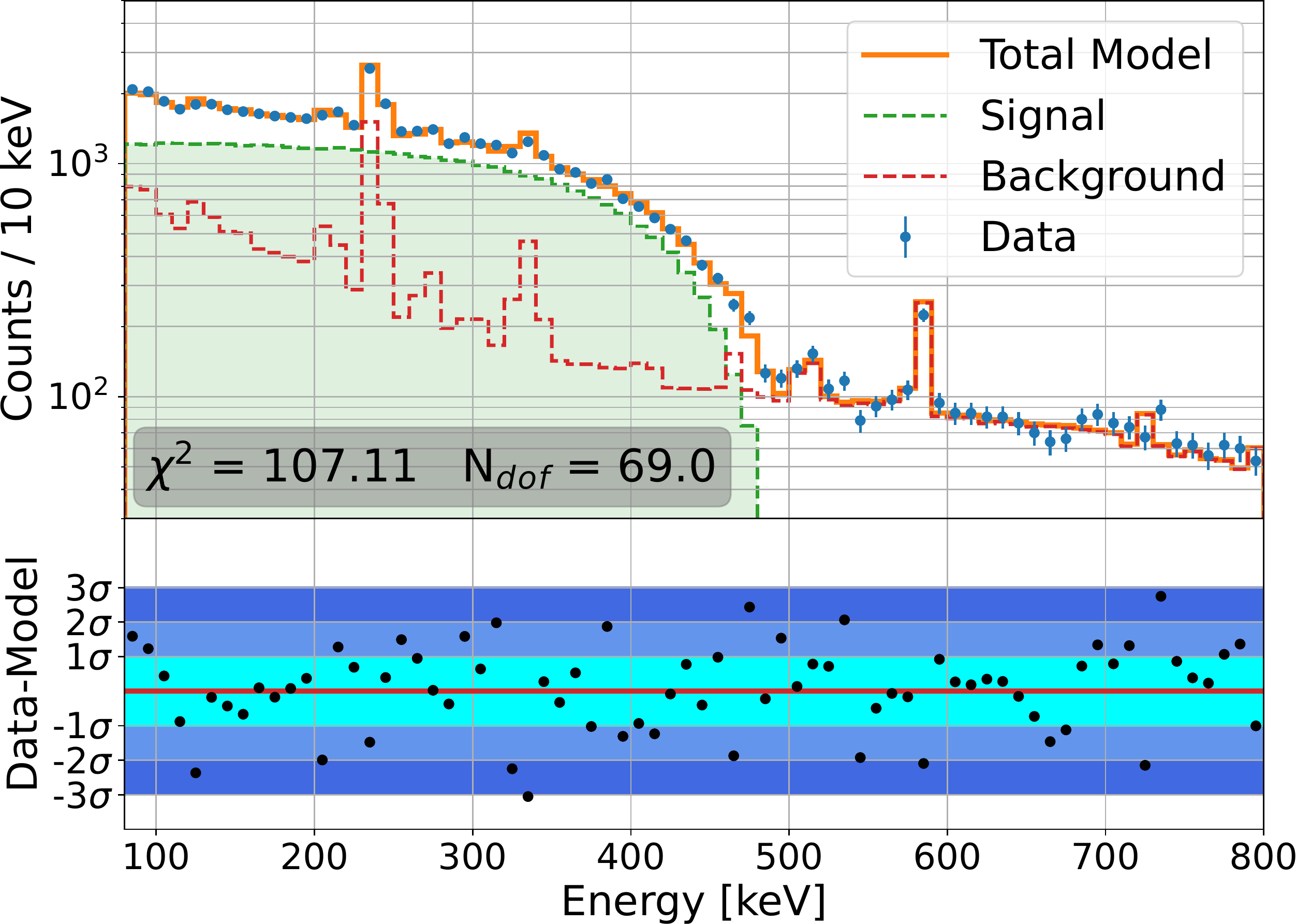}
  \caption{{\it Top.} Experimental spectrum (blue dots) and best fit result (orange solid line) obtained within the ISM model, which results to be the most suitable to describe the data in the current framework. The model resulting from the fit is a linear combination of the \ce{^{115}In} $\beta$-decay template spectrum (green dashed line), and the two contributions from the thorium calibration source (red dashed line). The $\chi^{2}$ and the number of degrees of freedom $\text{N}_{\text{dof}}$ are reported. {\it Bottom.} Fit residuals normalized to the statistical uncertainty.}
  \label{fig:bestFitExample}
 \end{figure}

\begin{figure*}[t]
    \centering 
    \includegraphics[width=2\columnwidth]{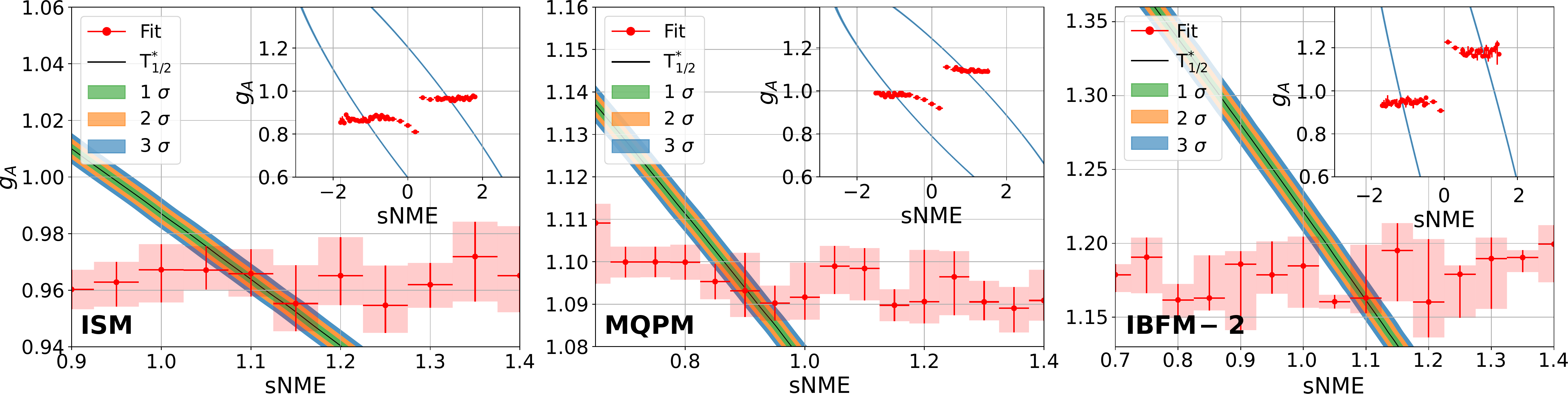}
    \caption{Identification of the optimal sNME values
    with the \emph{matched half-life fit} for ISM (left), MQPM (center) and IBFM-2 (right), respectively. The main plot reports as colored bands the half-life $T_{1/2}^{*}$ together with its uncertainties, and as red points the value of $g_A$ that best fit the data for a fixed value of sNME. The uncertainty on the latter is fixed by the template fine structure, while the one on $g_{A}$ is the [16, 84]\% quantile interval from the Bayesian fit. Each inset shows the half-life dependence on the other two parameters of the theory, together with the fit results, in a wider sNME interval.}
    \label{fig:Method1}
\end{figure*}

{\it Analysis Results.} We perform the data reconstruction by using two different fit methods.
The first method is the \emph{best fit}, which determines the configuration that best matches the data by letting both $g_{A}$ and sNME vary. For instance, Figure~\ref{fig:bestFitExample} depicts the data reconstruction through the best fit method achieved by using the template coming from the ISM model.
The second method, referred to as \emph{matched half-life fit}, tests the core of the sNME approach, namely the joint prediction of spectral shape and half-life of a forbidden $\beta$-decay. We vary the value of $g_{A}$ while fixing sNME, treating it as a free parameter of the model. We then select the sNME by comparing the fit result with the known half-life $T_{1/2}^{*}$ in the $(g_{A},\text{sNME})$ parameter space, where the trend of $T_{1/2}^{*}$ is predicted by the nuclear model being studied. The value of $T_{1/2}^{*}$ has been obtained as an average of previous measurements~\cite{PhysRevLett.129.232502, PhysRevC.19.1035, doi:10.1080/14786436208201861, PhysRev.122.1576}, weighted for their uncertainties, and is $T_{1/2}^{*}=(5.14 \pm 0.06) \times 10^{14}$\,yr.
This method is illustrated in Figure~\ref{fig:Method1} for the three nuclear models.

The results for the two fit methods in the three theoretical frameworks are summarized in Tab.~\ref{tab:results}, where both negative and positive solutions for sNME are reported for completeness.
For each combination of the fit method and model, the positive sNME solution is preferred based on the reduced chi-square $\chi^{2}_{red}$. 
When considering negative solutions, the resulting half-life is not compatible with $T^{*}_{1/2}$. Moreover, within the best fit method for negative values of sNME, the minimization process brings this parameter to its range limits, making the outcomes less reliable. In light of this, our discussion will focus on the positive sNME solutions. 

Considering the best fit method, we study the systematic effects due to the fit assumptions. As already mentioned, the half-life values exhibit perfect agreement when changing the nuclear model. Conversely, $g_A$ and sNME are strictly related to the approximations done within a specific theoretical framework, therefore we do not expect them to coincide. Moreover, we reiterate the best fit by assuming secular equilibrium in \ce{^{232}Th} decay chain contained in the calibration source. We also study the binning effect by changing the bin width to 20 keV and varying both the upper energy limit to 550 keV and 1000 keV and the lower energy threshold to 150 keV. We do not include a test with an energy threshold below 80 keV since we cannot have a satisfactory reconstruction of the background below this energy.
All the outcomes show values for $g_A$, sNME, and half-life completely compatible within 1$\sigma$ with the nominal ones reported in Tab.~\ref{tab:results}.

{\it Discussion.} Considering the best fit method, we consistently achieve a robust data reconstruction, obtaining a $\chi^{2}_{red}$ in the range [1.55, 1.66].
The signal-to-background ratio of the collected data limits the possibility of precisely determining $g_A$ and sNME simultaneously. The latter has a weaker impact on the spectral shape, therefore it is affected by a relatively high uncertainty, sometimes larger than 20\%. 

We observe a clear preference for positive sNME solutions, aligning closely with the CVC predictions. In particular, the experimental values are consistently around 30\% of the CVC ones (Tab.~\ref{tab:sNME_CVC}). Demonstrating a systematic preference for physical solutions near CVC values is crucial. This information significantly helps in selecting the correct spectral shape when it strongly depends on sNME value, as in some cases discovered in $\beta$-decay shape survey in Ref.~\cite{Ramalho2024}.

For the three models, we obtain different values of $g_{A}$, still, they all strongly reject the free-nucleon hypothesis with a significance of at least $4.7~\sigma$.
We can determine the half-life of the $^{115}$In $\beta$-decay with an accuracy of $\mathcal{O}(1\%)$ and all the obtained half-lives are fully compatible with each other.
Furthermore, these values are in agreement with $T_{1/2}^{*}$ within $1.4~\sigma$, regardless of the theoretical model.
However, the half-lives predicted by the models for the best fit parameters are $2.37\times10^{14}$\,yr (ISM), $8.52\times10^{13}$\,yr (MQPM) and $7.93\times10^{14}$\,yr (IBFM-2), far from the ones obtained with the fit.

It is therefore interesting to compare the best fit outcomes with those of the matched half-life fit, investigating how the predicted half-life match impacts the results. 
In the cases of ISM and IBFM-2, the fit quality mildly worsens and the physical parameters $g_{A}$ and $T_{1/2}$ are compatible, affirming the reliability and robustness of this method. 
By construction, the theoretical predictions on the half-lives agree with $T_{1/2}^{*}$ and are compatible within 1\,$\sigma$ with the measured half-life.
Conversely, the matched fit approach for the MQPM makes the model unable to describe the spectral shape.
Moreover, the resulting half-life in this case is not compatible within 2\,$\sigma$ with both theoretical predictions and all the other half-life determinations. 
This makes the joint prediction of spectral shape and half-life less reliable for this method.
   
The value of $g_A$ reported for $^{115}$In in Ref.~\cite{PhysRevLett.129.232502} are significantly smaller than the ones obtained in the current work. It seems that usage of the sNME degree of freedom not only improves the agreement between experimental and theoretical values of the half-life, but also shifts $g_A$ to bigger values. The same happens in $^{113}$Cd for MQPM when going from the analysis in Ref.~\cite{Bodenstein-Dresler:2018dwb} to the one in Ref.~\cite{KOSTENSALO2021136652}, while for ISM and IBFM-2 the two results are compatible.
The analysis based on the Spectral Moments Method in Ref.~\cite{Kostensalo:2023xzu} applies a technique somehow similar to the matched half-life fit of this work. Even if applied on \ce{^{113}Cd} data, the results quoted in terms of $g_A$ are very close to this work.

In summary, these findings deserve careful examination and further insights in future nuclear-model computations. It is crucial that such calculations include predictions of $\beta$-decay spectral shapes based on the preferred values of the sNME within the enhanced SSM framework. The present study is an important step towards possible systematic sNME preference schemes in this respect. At the same time, this work provides valuable insights into the evolution of the favoured values of the axial coupling when going from the SSM to the enhanced one.


\begin{table*}[t]
   \caption{Results for the two fit methods and the three considered nuclear models on the parameters of interest $g_{A}$, sNME and $T_{1/2}$. The reduced chi-square $\chi^{2}_{red}$ is also reported, quantifying the goodness of fit.}
   \setlength{\tabcolsep}{5pt}
     \begin{threeparttable}
     \begin{tabular}{l cccc|cccc}
            &\multicolumn{4}{c}{\it Positive solution}  &\multicolumn{4}{c}{\it Negative solution}\\[3pt]
      Model &$g_{A}$ & sNME & $T_{1/2}$ [$\times 10^{14}$\,yr] & $\chi^{2}_{red}$ & $g_{A}$ & sNME & $T_{1/2}$ [$\times 10^{14}$\,yr] & $\chi^{2}_{red}$\\
      \hline  \\[-6pt] {\it Best fit} \\ \cline{1-1} \\[-6pt]      
      ISM               & $0.964_{-0.006}^{+0.010}$  & $1.75_{-0.08}^{+0.13}$  & $5.26\pm 0.06$ & $1.55$  
                        & $0.774_{-0.042}^{+0.046}$  & $-5.43_{-0.22}^{+0.40}$\xspace$^{(*)}$ & $5.40\pm 0.07$ & $2.27$ \\[3pt]
      MQPM              & $1.104_{-0.017}^{+0.019}$  & $2.88_{-0.71}^{+0.49}$ & $5.26\pm 0.07$ & $1.65$
                         & $0.978_{-0.021}^{+0.022}$  & $-5.40_{-0.53}^{+0.38}$\xspace$^{(*)}$& $5.46\pm 0.07$ & $2.26$ \\[3pt]
      IBFM-2            & $1.172_{-0.017}^{+0.022}$  & $0.81_{-0.24}^{+0.52}$  & $5.25\pm 0.07$ & $1.66$ 
                        & $0.739_{-0.058}^{+0.069}$  & $-5.20_{-0.41}^{+0.63}$\xspace$^{(*)}$ & $5.40\pm 0.06$ & $1.97$\\[5pt]
      \hline  \\[-8pt] {\it Matched half-life} \\ \cline{1-1} \\[-6pt]
      ISM               & $0.965_{-0.010}^{+0.013}$  & $1.10\pm 0.03$ & $5.20\pm 0.07$ & $1.78$ 
                        & $0.869_{-0.004}^{+0.004}$  & $-1.15\pm 0.03$ & $5.50\pm 0.06$ & $2.94$\\[3pt]
      MQPM              & $1.093_{-0.007}^{+0.009}$  & $0.90\pm 0.03$ & $5.05\pm 0.06$ & 2.32  
                        & $0.992_{-0.004}^{+0.004}$  & $-1.00\pm 0.03$ & $5.64\pm 0.07$ & $3.22$\\[3pt]
      IBFM-2            & $1.163_{-0.010}^{+0.036}$  & $1.10\pm 0.03$ & $5.28\pm 0.06$ & $1.67$ 
                        & $0.958_{-0.015}^{+0.012}$  & $-1.15\pm 0.03$ & $5.46\pm 0.07$ & $2.28$\\[5pt]
     \end{tabular}
     \begin{tablenotes}
        \item[(*)] Posterior overlapping parameter boundaries.
     \end{tablenotes}
     \end{threeparttable}
    \label{tab:results}
   \end{table*}

\begin{acknowledgments}
This project has received funding from the European Union’s Horizon 2020 research and innovation program under the Marie Skłodowska–Curie grant agreement N.~101029688. This work was supported by the Academy of Finland, Grant Nos.~314733, 320062, and 345869. We thank the CUPID collaboration for sharing their cryogenic infrastructure, M. Guetti for the assistance in the cryogenic operations, M. Perego for his invaluable help in many tasks, the mechanical workshop of LNGS. This work makes use of the DIANA data analysis and APOLLO data acquisition software which has been developed by the CUORICINO, CUORE, LUCIFER and CUPID-0 collaborations.
\end{acknowledgments}
\bibliography{main}
\end{document}